\documentstyle[preprint,aps]{revtex}

\draft

\begin{document}
\preprint{NCL94-TP3}

\title{When q=0, The Forced Harmonic Oscillator Isn't.}

\author{J. W. Goodison\cite{email} and D. J. Toms\cite{emailtwo} }
\address{Physics Department, University of Newcastle upon Tyne,\\ Newcastle
upon
Tyne, NE1 7RU, U.K.}
\date{\today}

\maketitle

\begin{abstract}
  We consider the forced harmonic oscillator quantized according
to infinite statistics ( a special case of the `quon' algebra proposed by
Greenberg ). We show that in order for the statistics to be consistently
evolved
the forcing term must be identically zero for all time. Hence only the free
harmonic oscillator may be quantized according to infinite statistics.
\vskip 1in
\end{abstract}


\narrowtext

\section{Introduction}

  The idea of an interpolating particle statistics to the familiar Bose and
  Fermi cases, although exciting, is not new. Over fifty years ago Gentile
\cite{gent}
  proposed statistics in which at most $k$ particles could occupy a single
  state. Lately, much attention has been focussed onto the so-called `quon'
  algebra proposed by Greenberg \cite{greenPRD}

\begin{equation}
a_i a_j^{\dagger} - q a_j^{\dagger} a_i = \delta_{ij}
\end{equation}
 where $q$ is a real parameter and the $a_j^{\dagger}$ ( $a_i$ ) are creation
 (annihilation) operators. It can be seen that as $q$ ranges from +1 to -1, the
 above relation interpolates between the classical commutation and
anticommutation
 relations appropriate for Bosons and Fermions respectively.
 Much work has been done with this relation: Fivel \cite{fivel} and Zagier
\cite{zag}
 showed that for $|q| < 1$ the resulting Fock space is well defined and that
all
 states have a positive definite norm. Greenberg demonstrated that for free
fields
 the TCP theorem and clustering remain valid.

  Recently, we considered quons in curved spacetime \cite{gtoms}. In particular
we examined
  Parker's proof of the spin-statistics theorem. We found that for $|q| <1$ a
  contradiction arose on attempting to consistently evolve the particle
  statistics through the dynamically evolving universe. ( We also noted that
  by taking the physical limit in which the dynamical part of the evolution
  became constant, and demanding that the generalized commutation relations
  imposed remain continuous in this limit, the same contradiction should occur
  in flat spacetime ).

  In this work we examine this contradiction more closely for the special
  case of infinite statistics

\begin{equation}
a_i a_j^{\dagger} = \delta_{ij}
\end{equation}
 obtained by substituting $q=0$ in relation (1). ( We note that relation (2)
  was one of the original motivations for the study of quons \cite{greenPRL};
it
is also
  intrinsically interesting as it defines a quantum group - see for example
\cite{jim1,jim2,jim3,wor2,wor3} ). Greenberg
  remarked that studying $q=0$ is much easier than studying general $|q|<1$
  although the results are often qualitatively the same.

  The structure of this paper is as follows. In section two, we review our
previous work applying it specifically to the case of infinite statistics; in
particular we demonstrate that if a system is initially quantized according to
infinite statistics it remains so for all time. We then utilise this result in
section three where we consider the forced harmonic oscillator in detail. In
section four, we look at how this result applies in $n$ dimensions, and
possible
implications for more general systems.

\section{Review; No Infinite Statistics From Dynamics In Curved Spacetime}

  In this section we review our recent work and demonstrate that a
contradiction
  arises when we attempt to consistently evolve infinite statistics through
  a dynamically evolving universe.
  The central idea is to suppose that we have a time dependent spacetime
  which for times $t\leq t_1$ and $t\geq t_2$ is flat, and for $ t_1<t<t_2 $
may
be
  dynamic. As a specific example we could consider the spatially flat
  Robertson-Walker spacetime

\begin{equation}
ds^2 = dt^2 - R^2(t) \bigl( dx^2 + dy^2 + dz^2 \bigr)
\end{equation}
 where $R(t)=R_1$ for $t\leq t_1$ ( which we shall call the in-region ) and
 $R(t)=R_2$ for $t\geq t_2$ ( the out-region ).

  We now consider a real scalar field $\Phi (x)$ which is quantized according
to
  infinite statistics in the in-region i.e. the field operator can be
  expanded in terms of the creation and annihilation operators as
  operators as

\begin{equation}
\Phi (x) = \sum_i \bigl( F_{i}(x)a_{i} + F_{i}^*(x)a_{i}^{\dagger} \bigr).
\end{equation}
  where

\begin{equation}
a_i a_j^{\dagger} = \delta_{ij}
\end{equation}
and $\{F_i(x)\}$ is a complete set of positive energy solutions to the
Klein-Gordon
equation. A similar expansion for $\Phi (x)$ may be made in the out-region

\begin{equation}
\Phi (x) = \sum_i \bigl( G_{i}(x)b_{i} + G_{i}^*(x)b_{i}^{\dagger} \bigr)
\end{equation}
  where

\begin{equation}
b_i b_j^{\dagger} - q b_j^{\dagger} b_i = \delta_{ij}
\end{equation}
  and again $\{G_i (x) \}$ is a complete set of positive energy solutions to
the
  Klein-Gordon equation. We note that we allow for the possibility of a change
in
  particle statistics between the in and out-regions. We also assume that the
  quantum fields we are considering interact with a gravitational field of
arbitrary
  strength but are otherwise free.

  In general, $a_i \not= b_i$ due to particle creation in the expanding
universe.
  Since both sets $\{F_{i}(x)\}$ and $\{G_{i}(x)\}$ are assumed to be complete,
  we may expand one in terms of the other; for a spacetime of the form given in
(1)
  this leads to the following relation:

  \begin{equation}
  a_i =  \alpha_{i}b_i + \beta_{i}^{*} b_i^{\dagger}
  \end{equation}
  where $\alpha_i$ and $\beta_i$ are the ( diagonal ) Bogoliubov coefficients
  first used by Parker \cite{park5} to study particle creation in the dynamic
universe.
  Substituting (8) and its hermitian conjugate into (5); and using (7) gives

\begin{equation}
\delta_{ij} = | \alpha_i |^2 \delta_{ij} + \alpha_i \beta_j b_i b_j
+ \beta_i^* \alpha_j^* b_i^{\dagger} b_{j}^{\dagger}
+  \beta_i^* \beta_j b_i^{\dagger} b_j + q \alpha_i \alpha_j^* b_j^{\dagger}
b_i.
\end{equation}
    If we now take the expectation value of (9) with the out-region vacuum
state
$|0, {\rm out}>$ defined by

  \begin{equation}
  b_{i} | 0, {\rm out} > \equiv 0.
  \end{equation}
then we discover

   \begin{equation}
   1 = | \alpha_i |^2
   \end{equation}
  If we now substitute this into relation (9) we obtain

\begin{equation}
0 =  \alpha_i \beta_j b_i b_j
+ \beta_i^* \alpha_j^* b_i^{\dagger} b_{j}^{\dagger}
+  \beta_i^* \beta_j b_i^{\dagger} b_j + q \alpha_i \alpha_j^* b_j^{\dagger}
b_i
\end{equation}
       Because the operators which appear on the RHS of this equation are
independent
  of each other, the coefficients must vanish separately. In particular

\begin{mathletters}
\begin{equation}
\beta_i^* \beta_j = 0
\end{equation}
\begin{equation}
q \alpha_i \alpha_j^* = 0
\end{equation}
\end{mathletters}
  These must hold when $i=j$. Then we obtain

\begin{mathletters}
\begin{equation}
\beta_i = 0
\end{equation}
\begin{equation}
q=0
\end{equation}
\end{mathletters}
  We conclude that if a real scalar field $\Phi (x) $ is quantized according
  to infinite statistics in the in-region, then it remains quantized according
  to infinite statistics in the out-region. However, we find that $|\alpha_i| =
1$
  and $\beta_i = 0$ implying that positive energy solutions in the in-region
  remain positive energy solutions in the out-region - this corresponds to an
  absence of particle creation in the dynamic universe. This is in
contradiction
  to Parker's argument \cite{park1} that, for a spacetime of the form given in
(3), in general
  $\beta_i \not= 0$. ( A number of spacetimes leading to $\beta_i \not= 0$ are
known; for example see \cite{park6,bd} ).

\section{Toy Model: The Forced Harmonic Oscillator}

  In order to investigate further, we construct a toy model of the situation
  described in section two. We examine a single oscillator quantized according
to
  infinite statistics with a forcing term $f(t)$ which acts only for
  $t_1<t<t_2$. We take our ( time dependent ) Hamiltonian to be

\begin{equation}
H(t) = N(t) \omega + f(t) a(t) + f^*(t) a^{\dagger}(t)
\end{equation}
  where we have set $\hbar=1$ for convenience. $N(t)$ is the `physical'
  number operator satisfying

\begin{equation}
[ N(t), a(t) ] = - a(t)
\end{equation}
 ( Greenberg gives an explicit formula for $N(t)$ but we will not require that
here ). Our quantization scheme is

\begin{equation}
a(t) a^{\dagger}(t) = 1
\end{equation}
  which we assume to hold true for all times $t$ ( this is consistent with
  the propagation of infinite statistics as shown in section two ). For times
  $t\leq t_1$ and $t\geq t_2$, $H(t)$ becomes the free Hamiltonian.

  We now construct the Heisenberg equation of motion for the annihilation
  operator. This yields

\begin{eqnarray}
i{d a(t) \over dt} &&= [ a(t), H(t) ]
 \nonumber \\ &&= \omega a(t) + f^*(t) \bigl(
1 - a^{\dagger}(t) a(t) \bigr)
\end{eqnarray}
  This relation differs considerably from the usual forced harmonic oscillator
  relation ( see for example Merzbacher \cite{merz} ) due to the presence of
the
$f^*(t)a^{\dagger}(t)a(t)$ term.
  However, this is easily dealt with; we act with $a(t)$ from the left on both
  sides of relation (18); then using (17) we obtain

\begin{equation}
a(t) \biggl( i {d a(t) \over dt} - \omega a(t) \biggr) = 0
\end{equation}
  Since $a(t)$ is non-zero, we find that the solution to the above relation is
given
  by

\begin{equation}
a(t) = a \exp ( - i \omega t )
\end{equation}
  If we now substitute this relation and its hermitian conjugate back into
  our original equation of motion for $a(t)$, we discover

\begin{equation}
0 = f^*(t) \bigl( 1 - a^{\dagger} a \bigr)
\end{equation}
  We now take the expectation value of this relation with the vacuum state
$|0>$
  defined by

\begin{equation}
 a|0 > \equiv 0
\end{equation}
  and find that

\begin{equation}
f(t) = 0
\end{equation}

  Thus we conclude the harmonic oscillator quantized according to infinite
  statistics makes sense only as a free system.

\section{Discussion}

  The result we obtained in section three has been arrived at as follows; first
we showed that if a real scalar field $\Phi$ was originally quantized according
to infinite statistics then it would remains so for all time. The only
assumption we made was that the field interacted with the gravitational field.
We then used this result to show that $\Phi$ had to be the free field i.e. it
could not interact with the gravitational field - this is the source then of
our
contradiction. Hence we must originally assume that $\Phi$ does not interact
with
the gravitational field. A similar argument could be presented for a scalar
field interacting with an electromagnetic field which can also lead to particle
creation.

  We also note that since the number of spacetime dimensions does not enter
into
the calculation in section two, we would expect the same results to be obtained
in any number of spacetime dimensions. ( This is a special case of Chen's
\cite{chen} recent result. ) The result obtained at the end of section
three should also hold in any number of spacetime dimensions.

  It is also possible to wonder if the results of section three apply to
general
$|q|<1$.
Since the same problem described in section two arises for $|q|<1$,  it may
have
the same root
cause. If it is impossible to define the forced harmonic oscillator for general
$|q|<1$, then the future of quantum field theories based on quons is extremely
bleak.

\acknowledgements

   JWG would like to thank the S.E.R.C. for financial support. DJT would like
to
thank the Nuffield Foundation for their
   support.


\begin{references}
\bibitem[\clubsuit]{email} E-mail: j.w.goodison@ncl.ac.uk
\bibitem[\spadesuit]{emailtwo} E-mail: d.j.toms@ncl.ac.uk
\bibitem{gent}  G. Gentile, Nuovo Cimento {\bf 17} 493 (1940)
\bibitem{greenPRD} O. W. Greenberg, Phys. Rev. D {\bf 43} 4111 (1991).
\bibitem{fivel} D. I. Fivel, Phys. Rev. Lett. {\bf 65} 3361 (1990).
\bibitem{zag} D. Zagier, Commun. Math. Phys. {\bf 147} 199 (1992).
\bibitem{gtoms} J. W. Goodison and D. J. Toms, Phys. Rev. Lett. {\bf 71} 3240
(1993)
\bibitem{greenPRL} O. W. Greenberg, Phys. Rev. Lett. {\bf 64}, 705 (1990).
\bibitem{jim1} M. Jimbo, Lett. Math. Phys. {\bf 10} 63 (1985).
\bibitem{jim2} M. Jimbo, Lett. Math. Phys. {\bf 11} 247 (1986).
\bibitem{jim3} M. Jimbo, Commun. Math. Phys. {\bf 102} 537 (1987).
\bibitem{wor2} S. Woronowicz, Commun. Math. Phys. {\bf 111} 613 (1987).
\bibitem{wor3} S. Woronowicz, Invent. Math {\bf 93} 35 (1988).
\bibitem{park5} L. Parker, Ph.D. Thesis, Harvard (1965), unpublished.
\bibitem{park1} L. Parker, Phys. Rev {\bf 183} 1057 (1969).

\bibitem{park6} L. Parker in {\sl Asymptotic structure of space-time}, edited
by
F. Esposito and L. Witten
( Plenum, New York, 1977 ).
\bibitem{bd}  N. D. Birrell and P. C. W. Davies, {\sl Quantum fields in curved
space} ( Cambridge University
 Press, 1982 ).

\bibitem{merz} E. Merzbacher {\sl Quantum Mechanics} ( J. Wiley, New York, 1970
)
\bibitem{chen} C. C. Chen, National Chung-Hsing University Preprint (1993)


\end{references}
\end{document}